\documentclass[prb,aps,unsortedaddress,superscriptaddress,reprint,floatfix]{revtex4-1}

\usepackage[T1]{fontenc} 
\usepackage[utf8]{inputenc}
\usepackage[dvips]{graphicx}
\usepackage{amsfonts}
\usepackage{amsmath}
\usepackage{amssymb}
\usepackage{exscale}
\usepackage{color}	
\usepackage{braket}	
\usepackage{multirow}	
\usepackage{epstopdf}
\usepackage{microtype}
\usepackage{xspace}
\usepackage[ngerman,english]{babel}

\usepackage{letltxmacro}

\LetLtxMacro{\ORIGselectlanguage}{\selectlanguage}
\makeatletter
\DeclareRobustCommand{\selectlanguage}[1]{%
	\@ifundefined{alias@\string#1}
	{\ORIGselectlanguage{#1}}
	{\begingroup\edef\x{\endgroup
			\noexpand\ORIGselectlanguage{\@nameuse{alias@#1}}}\x}%
}
\newcommand{\definelanguagealias}[2]{%
	\@namedef{alias@#1}{#2}%
}
\makeatother

\definelanguagealias{en}{english}
\definelanguagealias{de}{ngerman}

\usepackage{moreverb} 
\immediate\write18{texcount -nc -inc -sum \jobname.tex > wordcount.tex}  

\immediate\write18{texcount -char -nc -freq \jobname.tex > charcount.tex} 

\usepackage[normalem]{ulem}	

\usepackage{siunitx}
\DeclareSIUnit{\oersted}{Oe}
\usepackage[version=3]{mhchem} 

\begin{document}
	
\title{Redox-controlled epitaxy and magnetism of oxide heterointerfaces: EuO/\ce{SrTiO3}}

\author{Patrick~Lömker}
\affiliation{Photon Science, Deutsches Elektronen Synchrotron, D-22607 Hamburg, Germany}		
\affiliation{Peter Grünberg Institut (PGI-6), Forschungszentrum Jülich GmbH, D-52428 Jülich, Germany}
\author{Martina~Müller}
\email{mart.mueller@fz-juelich.de}
\affiliation{Peter Grünberg Institut (PGI-6), Forschungszentrum Jülich GmbH, D-52428 Jülich, Germany}
\affiliation{Fakultät Physik, Technische Universität Dortmund, D-44221 Dortmund, Germany}

\date{\today}

\begin{abstract}
	We demonstrate a novel route to prepare thin films of the ferromagnetic insulator Europium monoxide. Key is a redox-controlled interface reaction between metallic Eu and the substrate \ce{SrTiO3} as the supplier of oxygen. The process allows tuning the electronic, magnetic and structural properties of the EuO films. Furthermore, we apply this technique to various oxidic substrates and demonstrate the universality and limits  of a redox-controlled EuO film synthesis.
\end{abstract}

\maketitle


Oxidic thin films, multilayers or heterostructures are becoming increasingly important in a wide range of cutting-edge applications~\cite{hwang_emergent_2012}. This includes novel field effect transistors, spin-, magneto-, and orbitronics, as well as topological oxide electronics~\cite{lorenz2016OxideElectronic2016, vazOxideSpinorbitronicsNew2018, uchidaTopologicalPropertiesFunctionalities2018, huang_emerging_2017}. 
Among the class of magnetic oxides, the ferromagnetic insulator Europium monoxide (EuO) represents one of the most intriguing functional materials, as it combines strong local $4f^7$ ferromagnetism with electrical insulation~\cite{muller_exchange_2009, mauger_magnetic_1986}. In that, EuO offers opportunity to probe quantum phenomena~\cite{prinz_quantum_2016} and can serve as a building block for a multitude of future spin-based applications such as magnetically gated two-dimensional electron systems with potential application of the inverse Edelstein effect~\cite{lomker_two-dimensional_2017, lesne_highly_2016,
song_observation_2016, bibe_towards_2012}. 

Oxide synthesis with atomic layer precision requires large experimental efforts, simultaneously varying the oxide properties introduces a further challenge. In the particular case of EuO, the adsorption-limited growth mode has become the standard approach for thin film growth~\cite{sutarto_epitaxial_2009}. This process is limited in two ways as it can only be applied to inert substrates and determining the appropriate deposition parameters is a complex task \cite{gerber_thermodynamic_2016}. The three parameters  substrate temperature, oxygen gas- and metal fluxes must be controlled simultaneously. 

The idea of the adsorption-limited growth process is to avoid over-oxidation of (metastable but ferromagnetic) EuO into (stable but paramagnetic) \ce{Eu3O4} and \ce{Eu2O3}. This is achieved by adjusting the flux ratio to Eu rich and taking advantage of the temperature-dependent re-evaporation, or 'distillation', of excess Eu metal \cite{steeneken_new_2002}.

In view of this laborious process, alternative growth schemes aim at simplifying experimental procedures. For instance, in a so-called 'topotractic' growth mode \ce{Eu2O3} is capped by a Ti metal top layer, which causes the oxidation of Ti to \ce{TiO2} and the reduction of \ce{Eu2O3} to stoichiometric EuO~\cite{mairoser_high-quality_2015}. The resulting EuO thickness was restricted to a few nanometers. This procedure limits the choice of over-layer material to Titanium and also requires a careful tuning of the \ce{Eu2O3}/Ti bilayer thickness for its complete reduction/oxidation. Moreover, the resulting \ce{TiO2} over-layer, a wide-band insulator, prevents the universal integration of EuO into functional heterostructures and transport devices.

For the advanced growth of oxide thin films and heterostructures, the substrate itself is usually not considered as an active part of the oxide growth process. However, most oxidic substrates do not act as 'passive' templates. Instead, they can cause undesirable changes in the properties of the over-layer, e.g. by  oxygen ion diffusion. To avoid chemical interactions on 'active' substrates a buffer layer is often applied in between substrate and film. Particular to EuO, previous reports found that adsorption-limited synthesis directly on \ce{SrTiO3} is not possible. The deposition of EuO is made possible by the application of a SrO or BaO buffer layer~\cite{miyazaki_fabrication_2012,iwata_preparation_2000}. The thermodynamics at the interface play an important role on the stoichiometry of the grown film~\cite{gerber_thermodynamic_2016}. 

In this work, we present a conceptionally reverse approach. A redox reaction between a metallic reactant, Eu, and an oxidic substrate, \ce{SrTiO3} (001), determines the stoichiometry of the growing Eu-oxide over-layer -- without addition of gaseous oxygen during synthesis. The temperature-dependence of the conductivity for ionic oxygen is used to control both the final thickness and the stoichiometry of the redox-grown EuO film. 

Our novel route for EuO synthesis allows to carefully control the chemical, magnetic and structural properties of the Eu oxides and thereby enhances the possibilities of oxide heterostructure preparation by a simple, effective bottom-up approach. Furthermore, we apply this technique to various oxidic substrates and demonstrate the universality of this redox-controlled oxide film synthesis. We believe the technique can be easily generalized to the design of other functional oxides thin films, interfaces and heterostructures.

\section{Experimental Details}

For the investigation of the redox growth of EuO on \ce{SrTiO3}(001) without additional oxygen gas we have developed a stepwise deposition method. As substrate we employed Nb:\ce{SrTiO3}(001) and pure Eu metal was evaporated from a Knudsen cell in an ultra high vacuum molecular beam epitaxy system. Further details are given in the supplementary. Deposition steps of pure Eu-metal alternate with \textit{in situ} structure analysis (reflection high-energy electron diffraction, RHEED, and low-energy electron diffraction, LEED) and chemical analysis (X-ray photoelectron spectroscopy, XPS). The stepwise deposition was carried out such, that the total deposition time was stopped to enable analysis at the times {$ t =  $\SIlist{1; 2; 5; 10; 20; 60}{\minute}}. Between two deposition steps, the sample is cooled from $ T_P $ to room temperature for analysis. This procedure is repeated for different substrate temperatures. In this way, a chemical and structural growth profile of europium oxide on \ce{SrTiO3} is obtained depending on the deposition time $ t $ and  temperature $ T_P $ as determined from the pyrometer (see supplementary).

\textit{In situ} XPS is conducted on Ti $ 2p $ and Eu $ 3d $ core-levels with a PHOIBOS-100 hemispherical energy analyzer using Al~K$_{\alpha}$ radiation from an X-Ray anode (SPECS). 

We characterize the structure \textit{ex situ} with X-ray diffraction (XRD), X-ray reflection (XRR) and reciprocal space mapping (RSM). Magnetic properties are analyzed with vibrating sample magnetometry and a magnetic property measurement system (See supplementary).

\section{Results}


\begin{figure}[!tb]
	\begin{center}
		\includegraphics[clip,width=0.47\textwidth]{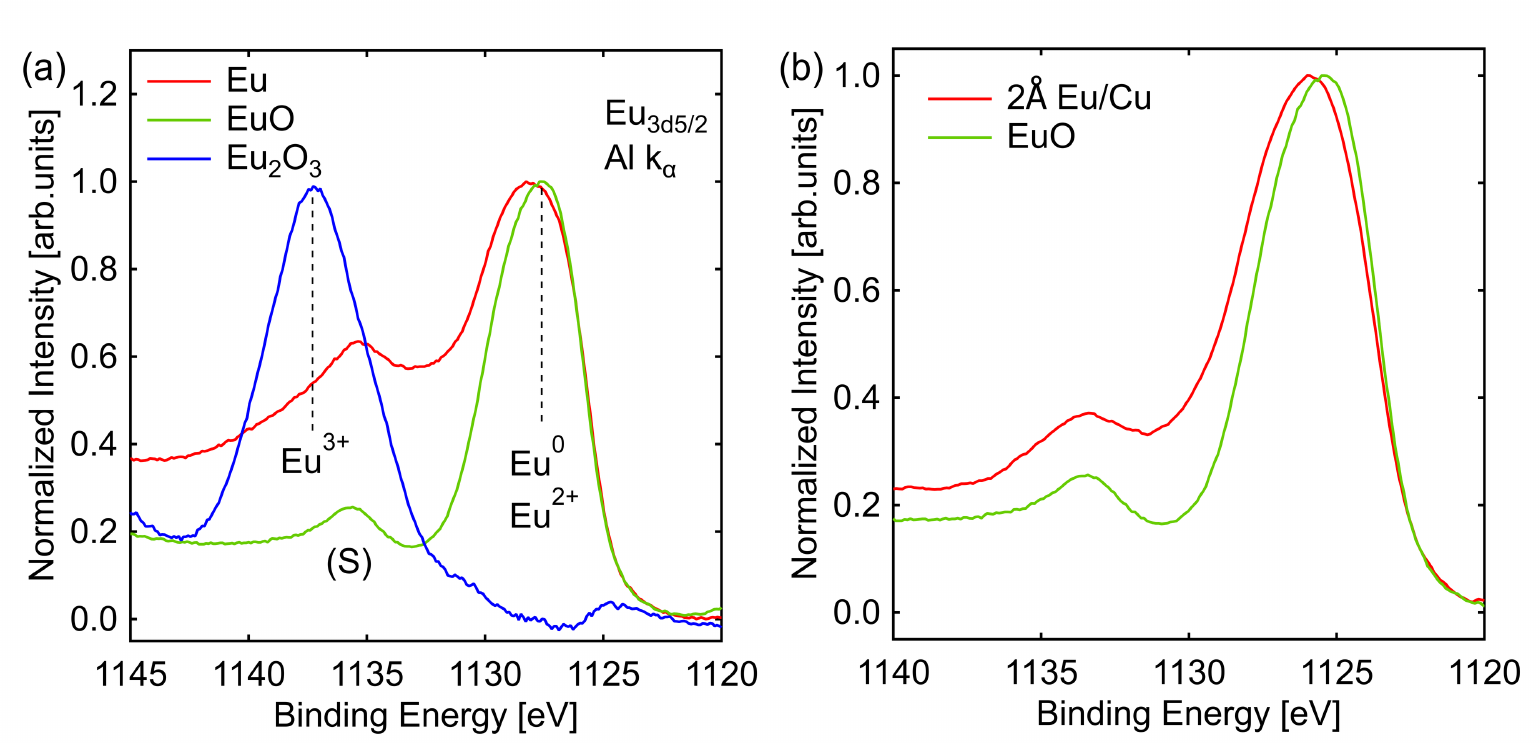}
		\includegraphics[clip, width=0.47\textwidth]{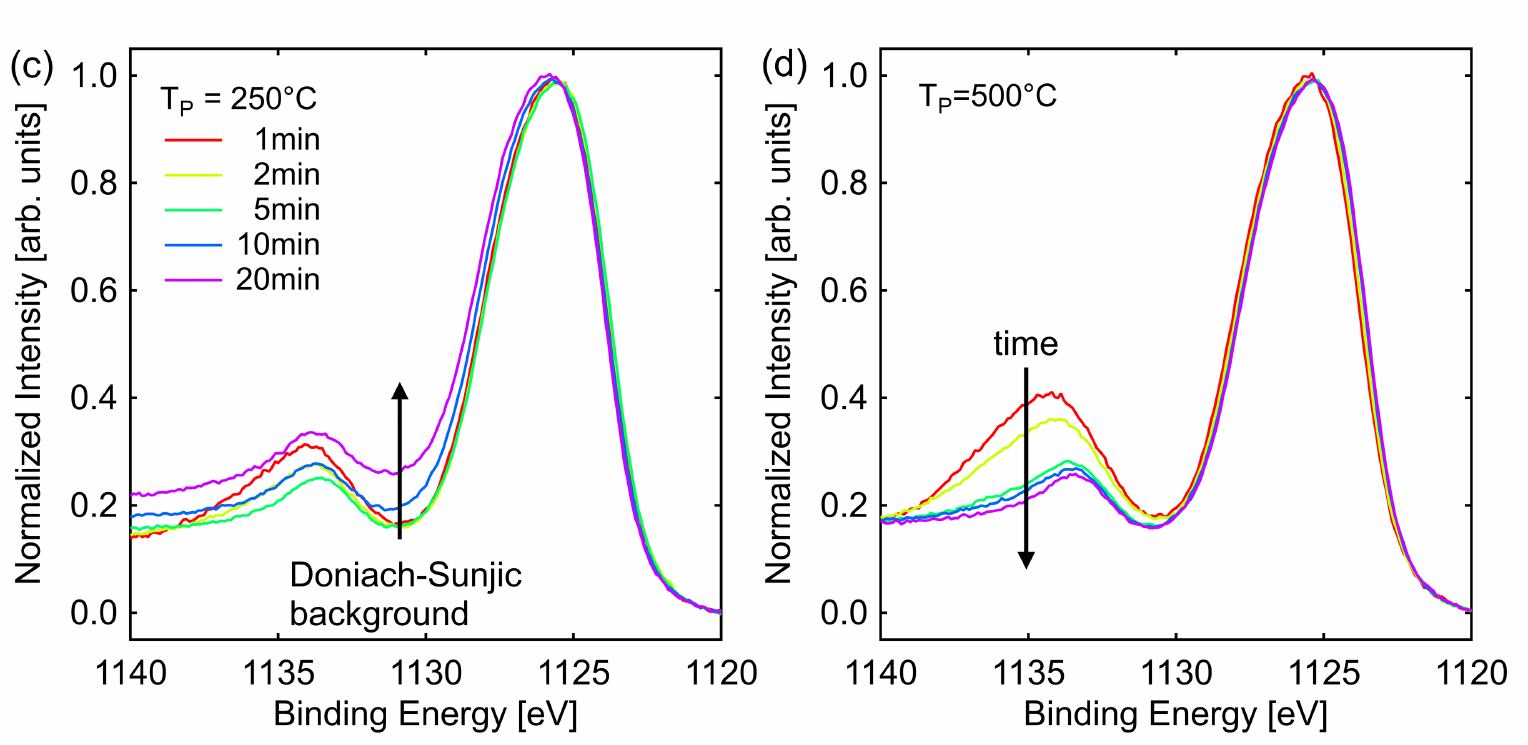}
	\end{center}
	\caption{\footnotesize{
			Top: (a) Reference spectra of  metallic, di- and trivalent Eu, EuO and \ce{Eu2O3} films respectively, obtained on the Eu~$ 3d_{5/2} $ core-level. (b) Comparison of stoichiometric EuO and $ d \approx \SI{2}{\angstrom} $ Eu metal on a metallic substrate. 
			Bottom: XPS analysis of the Eu $ 3d_{5/2} $   for a stepwise EuO redox growth below (c)  and above  (d) the re-evaporation temperature of Eu. For $ t\geq \SI{5}{min} $ metal inclusions are observed, while at elevated temperature a decreasing \ce{Eu^3+} content is present.
		}
	}
	\label{fig:fig1}
\end{figure}

Our study combines the redox-driven EuO synthesis with a chemical and structural\textit{ in situ} (XPS, LEED, RHEED) and magnetic and structural \textit{ex situ} (SQUID, XRD, XRR, RSM) analysis.

First, we acquire reference spectra of the Eu~$3d _{5/2} $ core-level (Fig.~\ref{fig:fig1}(a)) as a function of the valence of the Eu cations. As reference systems, we prepare films of pure phases ($ d = \SI{10}{nm} $) of \ce{Eu^0} (Eu metal), \ce{Eu^2+} (EuO) and \ce{Eu^3+} (\ce{Eu2O3}, see supplementary). The Eu $ 3d_{5/2} $ peak of divalent \ce{Eu^2+} is located at $E_B = \SI{1125}{\eV}$~\cite{caspers_chemical_2011, gerber_thermodynamic_2016, cho_origin_1995}.  This peak is accompanied by a satellite peak (S) at higher binding energy, which is part of the multiplet of the $3d^{9}$ $4f^{7}$ final state~\cite{cho_origin_1995}. Eu metal is observed at the same binding energy. In order to clearly distinguish the reference spectra of metallic Eu from the divalent state, we use the Doniach-Sunjic inelastic background, which is only present in metallic samples.  At $ E_{B} = \SI{1135}{eV} $ the Eu $ 3d_{5/2} $ peak is detected for trivalent \ce{Eu^3+}. At $ E_{B} = \SI{1125}{eV} $ an X-ray satellite from the trivalent Eu $ 3d_{5/2} $ is observed as a consequence of using non-monochromatized Al X-rays.

The sensitivity of the Doniach-Sunjic background shape as a measure for metallic \ce{Eu^0} is demonstrated by comparing a stoichiometric EuO film on yttria-stabilized zirconia (YSZ(001)) with $ d \approx \SI{2}{\angstrom} $ Eu metal deposited on a Cu(001) single crystal. The Doniach-Sunjic inelastic background for this sample is clearly observed (Fig.\ref{fig:fig1}(b)), showing that even mono-layers of Eu metal can be detected by this principle.

Knowing the reference spectra, we analyze in detail the redox-growth of EuO/\ce{SrTiO3} at two exemplary temperatures, one being below the re-evaporation temperature of the distillation process and one above that temperature. For the first five minutes of deposition at $ T_P = \SI{250}{\degreeCelsius} $ mainly intensity from \ce{Eu^2+} species are observed (Fig. \ref{fig:fig1} (c)). Less than 10\% of \ce{Eu^3+} is detected and the \ce{Eu^3+} content reduces over time. However, for $t>$\SI{5}{\minute} we observe Doniach-Sunjic inelastic background, indicating Eu metal inclusions in the film.  Further deposition leads to an increase in background and indicates \ce{Eu^0}. We conclude, that only a small amount of EuO is formed at this temperature for the initial growth. For extended growth the Eu metal inclusions turn the stoichiometry to Eu-rich.

Next, we study a stepwise Eu deposition at $T_P=$\SI{500}{\celsius} (Fig.~\ref{fig:fig1}(d)).  
For $t=$\SI{1}{\minute}, we observe a spectrum with a mixture of Eu$^{2+}$ (85\% )and Eu$^{3+}$ (15\%) components, while contributions of Eu metal are absent. For continued growth the spectral weight from Eu$^{3+}$ decreases (1\% at $t=$\SI{20}{\minute}). Already for $t>$\SI{5}{\minute}, we find that the XPS spectra of adsorbed Eu-metal are nearly indistinguishable from stoichiometric EuO reference data (compare Fig.\ref{fig:fig1}(a)). This demonstrates that the Eu metal is oxidized into a Eu$^{2+}$ rich (Eu$^{2+}$,Eu$^{3+}$) mixture at the \ce{SrTiO3} interface. For extended growth only stoichiometric EuO (Eu$^{2+}$) is observed.

\begin{figure}[!tb]
	\begin{center}
		\includegraphics[clip,width=0.47\textwidth]{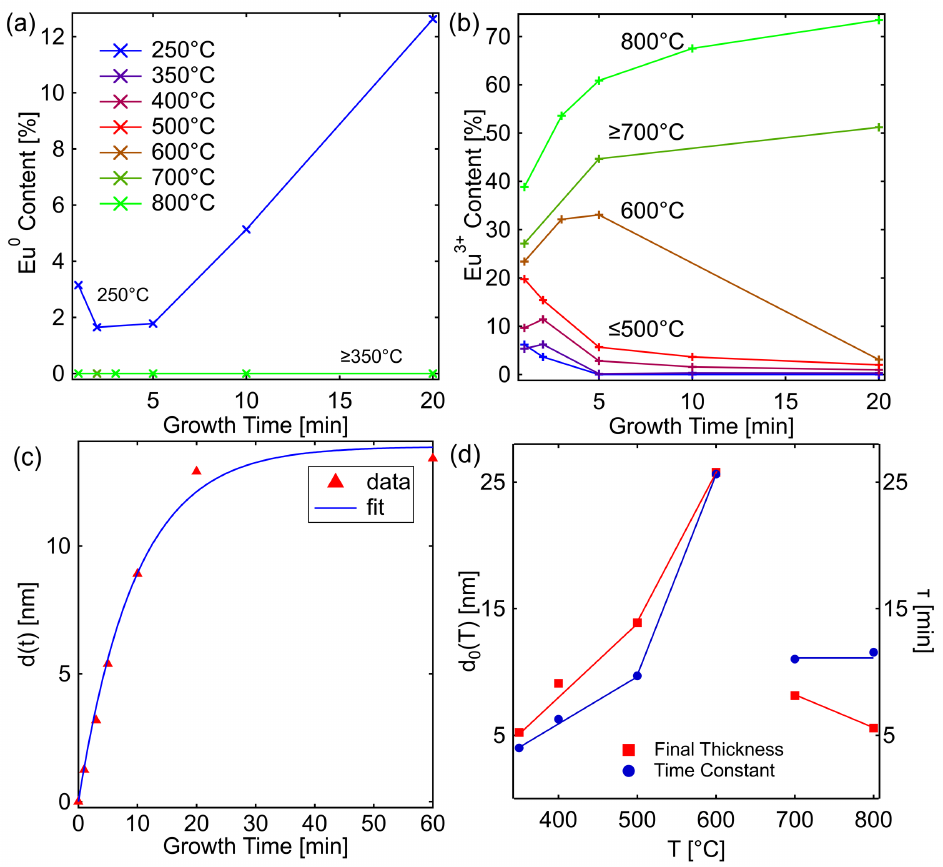}
		\includegraphics[width=0.47\textwidth]{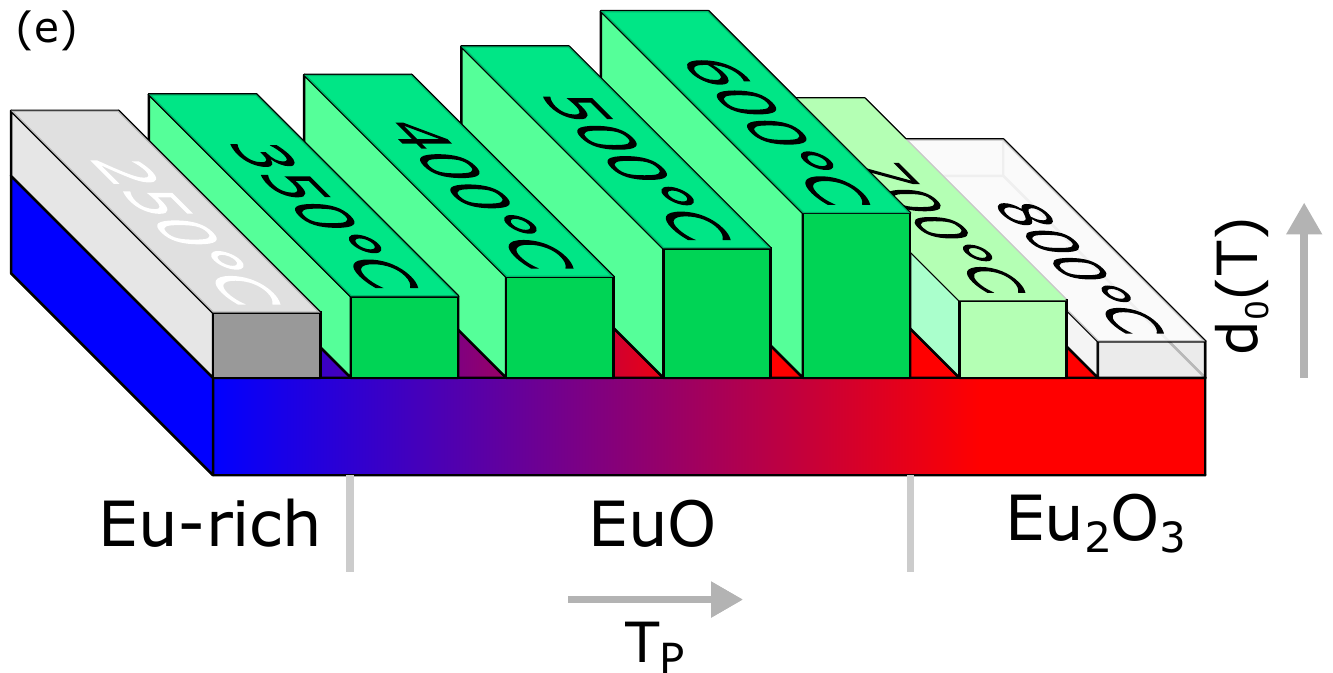}
	\end{center}
	\caption{\footnotesize{
			Chemical quantification of the Eu $ 3d_{5/2} $ core-level for metallic Eu$ ^{0} $  (a) and over-oxidized Eu$ ^{3+} $ (b) content as function of $ t $ and $ T_{P} $. (c) Exponential thickness dependence exemplary shown for a sample grown at $ T_P = \SI{500}{\degreeCelsius} $. (d) A reduction in $ d_0 $ and $ \tau $ is observed at $ t \geq \SI{700}{\degreeCelsius} $ at the same temperature where a \ce{Eu^3+} rich growth is observed. (e) We summarize the findings in a block diagram depicting the overall trend of the resulting film thickness and stoichiometry as function of $ T_P $.
		}
	}
	\label{fig:fig2}
\end{figure}

In the following, the stoichiometry of the grown film is quantified by fitting a linear combination of the reference spectra to the observed Eu \textit{3d}$ _{5/2} $ spectra.  First, we discuss the Eu metal content of the grown films (Fig.~\ref{fig:fig2}(a)). 
In addition to the findings presented for $ T_{P} = \SI{250}{\degreeCelsius} $, we find for $ T_P\geq\SI{350}{\celsius}$ no \ce{Eu^0} metal in the spectra. We conclude that now Eu re-evaporation is a dominant process, which is in line with previous reports utilizing the adsorption limited growth mode on YSZ~\cite{steeneken_new_2002}.

We observe an exponential decay of the \ce{Eu^{3+}} fraction with $t$ for films prepared at {$T_P= \,$\SI{250}{\celsius},} \SI{350}{\celsius}, \SI{400}{\celsius} and \SI{500}{\celsius}  (Fig.~\ref{fig:fig2}(b)).  The Beer-Lambert law predicts an exponential decay of the intensity for the case of a buried layer with the growth over-layer thickness. Therefore, we argue that only the interface is responsible for the \ce{Eu^{3+}} formation. It is also noted, that the \ce{Eu^3+} content is higher at $ t = \SI{1}{\minute} $ for increasing $ T_{P} $. Yet, the total amount of \ce{Eu^{3+}} is below one ML of \ce{Eu2O3}, which is a negligible amount for extended film growth ($ T_{P} = \SI{500}{\degreeCelsius} $).
For $ T_P=\SI{600}{\celsius}$,  the \ce{Eu^{3+}} content increases with $t$ in the initial growth phase ($ t\leq \SI{5}{\minute}$). For the extended growth ($ t \geq \SI{5}{\minute} $) the content of \ce{Eu^3+} decreases again and at $ t =\SI{20}{\minute} $ only \ce{Eu^2+} is found in the spectrum. Increasing the temperature further to \SI{700}{\celsius}, and \SI{800}{\celsius}, the content of \ce{Eu^{3+}} increases monotonically with time and no EuO growth phase is observed. Therefore, redox growth for stoichiometric EuO on \ce{SrTiO3} is possible only in the range {$ T_P =\, $\SIrange{350}{600}{\celsius}}.

\subsection{Thickness dependence}
Exemplary shown for a sample grown at $ T = \SI{500}{\degreeCelsius} ,$ we find the temporal dependence of the thickness,  $ d(t) $ (see supplementary), to follow an exponential law of the form $ d(t) = d_0 ( 1 - \exp( - t / \tau ) ) $ (Fig.~\ref{fig:fig2}(c)). Here $ d_0 $ is the final thickness and $ \tau $ is the time constant. We apply this model to all samples (Fig~\ref{fig:fig2}(d)) and find that both $ d_0 $ and $ \tau $ increase monotonically for $ T \leq \SI{600}{\degreeCelsius} $. Above this temperature a significant reduction is observed for both $ d_0 $ and $ \tau $, simultaneous to the transition to \ce{Eu^3+}-rich growth as observed in the chemical analysis.

We find that EuO rich films (i.e. $ \SI{350}{\degreeCelsius} \leq T_{P} \leq \SI{600}{\degreeCelsius}$) can be grown with up to $ d=\SI{25}{nm} $ and $\tau=$\SI{25}{\minute}. The fact that the time constants are in the range of many minutes allows a precise control of the film thickness by stopping the growth at a suitable time. The resulting chemical composition and thickness are compiled in a block diagram as a function of temperature (Fig.~\ref{fig:fig2}(e)).

\subsection{Structural profile}

\begin{figure}[!tb]
	\begin{center}
		\includegraphics[clip, width=0.47\textwidth]{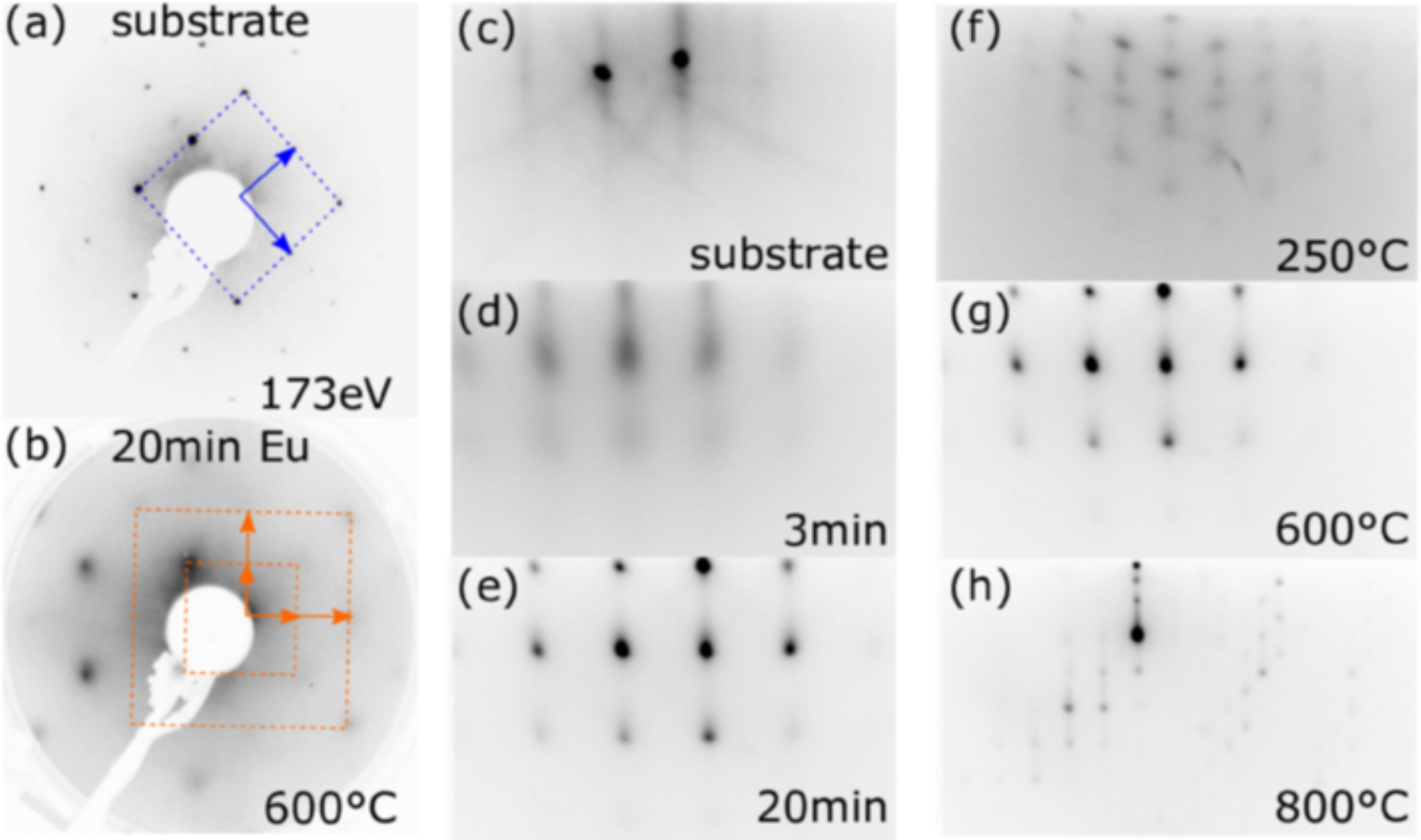}
	\end{center}
	\caption{\footnotesize{
			\textit{In situ} structure determination by LEED of the \ce{SrTiO3}(001) substrate (a) and the redox-grown EuO film (b). We observe a \ang{45} rotation of the EuO basis. (c-e) RHEED as a function of time shows a weak island growth mode and (f-h) structural transitions as a function of $ T_P $  from \ce{Eu^0} to \ce{Eu^2+} and \ce{Eu^3+} respectively. The RHEED beam is parallel to \ce{SrTiO3}(110) using \SI{20}{keV} electrons.
		}
	}
	\label{fig:fig3}
\end{figure}


For the \textit{in situ} stuctural analysis we first present LEED (Fig.~\ref{fig:fig3}(a)) for the \ce{SrTiO3} substrate. We observe clear and sharp reflexes. The LEED pattern is four-fold symmetric, reflecting the symmetry of the perovskite lattice. Indicated by blue arrows is the basis of the reciprocal lattice. The EuO film ($ T_P = \SI{600}{\degreeCelsius} $, Fig.~\ref{fig:fig3}(b)) is well-ordered and exhibits LEED reflexes, with a basis rotated by \ang{45} to that of \ce{SrTiO3}, the epitaxial relationship is {EuO(110)/\ce{SrTiO3}(100)}. The LEED reflexes are wider indicating a small degree of mosaicity.

In Fig.~\ref{fig:fig3}(c-e) we show RHEED of the substrate, Eu deposition for $ t = \SI{3}{\minute} $ and $ t = \SI{20}{\minute} $ at $ T_{P} = \SI{600}{\degreeCelsius}$. The substrate shows RHEED streaks and two sharp spots on the Laue-circle, indicating a flat and well oriented substrate. At three minutes Eu deposition we observe RHEED streaks and a weak transmission pattern. For continued growth the transmission pattern dominates the RHEED reflexes, indicating a island type growth mode.

The RHEED reflexes as a function of $ T_{P} = $ \SIlist{250;600;800}{\degreeCelsius} are shown in Fig.~\ref{fig:fig3}(f-h). For Eu-rich growth at \SI{250}{\degreeCelsius} we observe the hexagonal RHEED reflexes of the Eu-metal crystal. At elevated temperature the expected EuO related reflexes dominate the RHEED pattern, as described above. Finally at $ T_{P} = \SI{800}{\degreeCelsius} $ the structure shows the much bigger unit cell, and therefore shorter distances between RHEED reflexes, of the cubic \ce{Eu2O3} crystal and a complex transmission pattern.

In conclusion, we observe clear structural transitions that are perfectly in line with the results from the chemical analysis by XPS. The EuO films grow by a redox-driven process epitaxially and are, except for a small surface roughness, flat. 

\begin{figure}[!tb]
	\begin{center}
		\includegraphics[clip, width=0.47\textwidth]{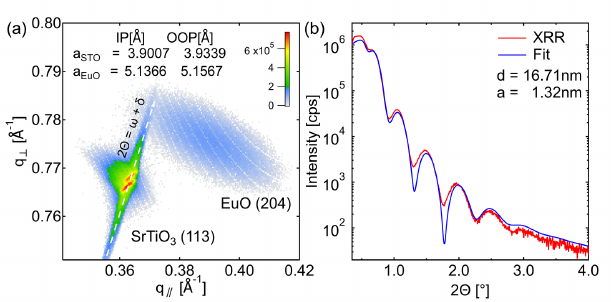}
	\end{center}
	\caption{\footnotesize{
			\textit{Ex situ} X-ray analysis of a sample grown at \SI{600}{\degreeCelsius}. The reciprocal space map (a) shows the {\ce{SrTiO3} (113)} and{ EuO (204)} peak in close proximity, a consequence of the epitaxial relation of {EuO(110)/\ce{SrTiO3}(100)}. The X-ray reflectometry  shows that the film has a low roughness of \SI{1.32}{nm} (b).
		}
	}
	\label{fig:fig4}
\end{figure}

In Fig.~\ref{fig:fig4}(a), we show the \textit{ex situ} X-ray analysis, starting with a RSM of a sample grown at $ T_{P} = \SI{600}{\degreeCelsius} $. The Bragg peaks from the \ce{SrTiO3}(113) and the EuO (204) peak are shown. The simultaneous observation of both reflexes is in line with the expected rotation of both lattices {EuO(110)$ \parallel $\ce{SrTiO3}(100)}, since the reflexes would otherwise be observed at an angle of \SI{45}{\degree}. The EuO(204) reflex has a rocking curve width of $ \delta\omega = \SI{1.1}{\degree} $ in line with the widened LEED reflexes. We calculate the lattice parameters for the in-plane and out-of-plane lattice constants and find that both \ce{SrTiO3} and EuO ($ a_{\ce{SrTiO3}} = \SI{3.901}{\angstrom} $ and $ a_{\ce{EuO}} = \SI{5.14}{\angstrom} $) are close to the literature values ($ a_{\ce{SrTiO3}} = \SI{3.905}{\angstrom} $ and $ a_{\ce{EuO}} = \SI{5.14}{\angstrom} $ \cite{karlsruhe_inorganic_2017}).

In Fig.~\ref{fig:fig4} (b), we depict the corresponding XRR curve and a fit to the measured data points using the Parrat formalism. The grown film exhibits a total thickness of $ d = \SI{16.7}{nm} $, while the roughness is $ a = \SI{1.3}{nm} $. Hence we conclude, that the redox-growth process produces well oriented, i.e. \SI{45}{\degree} rotated, EuO films on \ce{SrTiO3} with a small degree of roughness.

\subsection{Magnetic Properties}
\begin{figure}[!tb]
	\begin{center}
		\includegraphics[clip, width=0.47\textwidth]{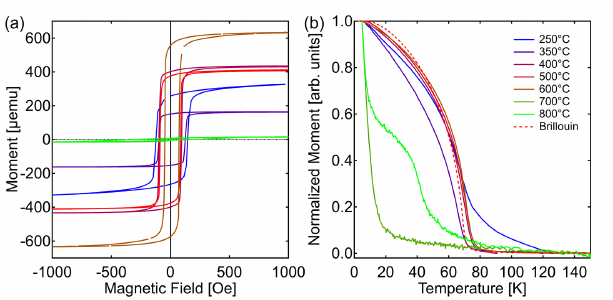}
	\end{center}
	\caption{\footnotesize{
			Magnetic properties of redox-grown \ce{Eu}-oxide thin films on \ce{SrTiO3} as (a) function of applied magnetic field and (b) temperature. EuO rich films, as found in the chemical analysis, exhibit a hysteresis, while high \ce{Eu^{3+}} contents lead to paramagnetism. The inclusion of \ce{Eu^{0}} as seen at $ T_{P} = \SI{250}{\degreeCelsius} $ causes a magnetic tail in the temperature dependence.
			}
	}
	\label{fig:fig5}
\end{figure}

The magnetic properties of samples grown at {$ T_{P} = $~\SIrange{250}{800}{\degreeCelsius}} are depicted in Fig~\ref{fig:fig5}(a) as hysteresis loops and as function of temperature $M(T)$ in Fig.~\ref{fig:fig5}(b). Since EuO is the only ferromagnetic component in the stack, it is expected, that the saturation magnetization $ M_S $ is proportional to the EuO thickness. Eu metal and \ce{Eu2O3} exhibit small paramagnetic moments only. In the chemical analysis we showed that the EuO content and the thickness both depend on $ T_{P} $ in a complex manner. Consequently, the magnetic analysis cannot be expected to follow a simple linear temperature dependence.

All $M(H)$ curves of $ T_{P} = $\SIlist{350;400;500;600}{\degreeCelsius} increase monotonically in $ M_{S} $, while the coercive field, $ H_{C} $, decreases simultaneously. Unordered and thin films are generally considered to cause an increase of the coercive field~\cite{muller_thickness_2009}. This finding suggests, that the amount of ferromagnetic EuO increases as a function of the temperature and forms a more ordered lattice at higher temperatures.

The samples grown at $ T_{P} = $\SIlist{700;800}{\degreeCelsius} exhibit no hysteresis, which is in line with the chemical analysis that reveals a \ce{Eu2O3}-rich (and therefore paramagnetic) composition.

The temperature dependence of the normalized magnetic moment is shown in Fig~\ref{fig:fig5}(b). We find for the samples grown at $ T_{P} = $\,\SIrange{350}{600}{\degreeCelsius} a Brillouin-like shape that closely follows a simulation with a Curie temperature of $ T_C = \SI{69}{K} $, the literature value for EuO \cite{matthias_ferromagnetic_1961}. Significant deviations are observed for samples grown at high temperature, where a paramagnetic behavior is measured and  for $ T_{P} = \SI{250}{\degreeCelsius} $, where a pronounced metallic tail indicates the presence of Eu metal ions included in a EuO film~\cite{suitsAnnealingStudyEuO1971, altendorf_oxygen_2011}.

\section{Discussion}

Unlike the classical EuO synthesis processes -- for which oxygen gas is supplied during a reactive MBE growth process -- here the oxide substrate itself acts as the supplier of oxygen: The \ce{SrTiO3} substrate is reduced by the presence of Eu while the reactant oxidizes to EuO, \ce{Eu3O4} or \ce{Eu2O3}. These processes can be assessed with an Ellingham analysis (see supplementary). Indeed, we find that at equilibrium the most likely formed oxide is \ce{Eu2O3}. However, we observe the formation of EuO in the intermediate temperature range. We therefore describe the complex redox growth of EuO/\ce{SrTiO3} as an interplay of three factors: (i) the kinetics of the oxygen anion reservoir from the substrate, (ii) the kinetics of the Eu metal on the surface of \ce{SrTiO3} (and its concomitant re-evaporation) and (iii) the thermodynamics of the interface reactions.


As seen in Fig.~\ref{fig:fig2}(c) the thickness of a redox-grown Eu oxide film is limited. This is surprising, as a normal diffusion type growth would have lead to a $ d \propto \sqrt{t} $ type behavior~\cite{einsteinUberMolekularkinetischenTheorie1905}. We explain the thickness limit as a consequence of the insulating properties of the Eu oxides: The \ce{O^{2-}} ions are charged and have to cross an insulating film of already grown Eu-oxides. This is well described in the context of a Mott-Cabrera type growth mode as was also found for the oxidation of Fe~\cite{cabrera_theory_1949, kruger_room_1964}. In the Mott-Cabrera type growth, the potential that the \ce{O^{2-}} ion is submitted to depends on the thickness of the film and its ion related resistivity. This explains the sudden reduction of $ d_0 $ and $ \tau $ at the transition to \ce{Eu^3+} rich growth in Fig.~\ref{fig:fig2}(d), as the resistance is larger for the higher Eu oxides. Also, the re-evaporation of Eu from the surface can be expected to be much faster.

\section{Outlook}
\begin{figure}[!tb]
	\begin{center}
		\includegraphics[clip, width=0.47\textwidth]{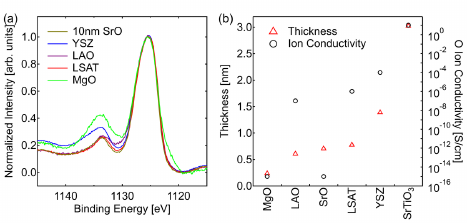}
	\end{center}
	\caption{\footnotesize{
			 (a) XPS analysis of the Eu$ 3d_{5/2} $ core-level for redox-grown Eu oxides, (b) calculated thickness $ d $ and oxygen ion conductivity $ \sigma $ for a selection of oxides grown at $ T_{P} = \SI{500}{\degreeCelsius} $ and $ t = \SI{5}{\minute} $. A qualitative correlation between $ d $ and $ \sigma  $ is observed.
		}
	}
	\label{fig:fig6}
\end{figure}

In order to explore the redox-driven growth of EuO more generally, we study the initial growth ($ t = \SI{5}{\minute} $) at elevated temperatures ($ T_{P} = \SI{500}{\degreeCelsius} $) of Eu(O) on YSZ, \ce{(LaAlO3){0.3}(Sr2TaAlO6){0.7}} (LSAT),  \SI{10}{nm} SrO grown on \ce{SrTiO3}, \ce{LaAlO3} (LAO),  and MgO and compare it to the previous results of \ce{SrTiO3}.

Again we study the Eu \textit{3d}$ _{5/2} $ core-level, as shown in Fig.~\ref{fig:fig6}(a).  In the case for MgO and LAO we observe \ce{Eu^0} in the film even at this high $ T_P $. For SrO and LSAT, we observe only \ce{Eu^2+}, whereas for MgO and YSZ we even observe some \ce{Eu^3+}. This is surprising, as preparations on YSZ are often reported in adsorption limited growth conditions and the interfacial over-oxidation is not mentioned \cite{sutarto_epitaxial_2009}.


We obtain the film thickness, $ d $, (see supplementary) for all redox-grown films and compare them to estimates for the ionic conductivities of the respective oxide substrate~\cite{skinnerOxygenIonConductors2003,gunkelInfluenceChargeCompensation2012,rudolphUeberLeitungsmechanismusOxydischer1959,nguyenEffectOxygenVacancy2000,mitoffElectronicIonicConductivity1962,moosDefectChemistryDonorDoped2005}. We find, that $ d $ qualitatively scales with the ionic conductivities of the underlying substrate (Fig.~\ref{fig:fig6}(b)). We attribute the high discrepancy between SrO conductivity and $ d $ of redox-grown EuO to the fact, that the SrO thin film grown on a \ce{SrTiO3} will have a higher oxygen mobility than a bulk crystal, due to crystal defects.

From this analysis we are able to differentiate substrates into active substrates, with a relevant redox process and passive substrates, where additional oxygen needs to be supplied. We find that \ce{SrTiO3} is the most active substrate for redox-growth. LAO, SrO thin films, LSAT and YSZ might pose as suitable templates for thin films and are expected to lead to thinner EuO over-layers than would be expected for \ce{SrTiO3}, while MgO and LAO act as passive substrates. 

\section{Conclusions}
In summary, we report a novel route for the synthesis of stoichiometric and single-crystalline EuO films by supplying no gaseous oxygen. Instead, utilizing the oxidic substrate as source of oxygen is the key to a reliable and simplified preparation scheme. We have identified the parameter window in which
EuO can be grown on \ce{SrTiO3}(001) and reduce the complexity of the typically applied distillation growth mode. 

The prepared films show the expected chemical, structural and magnetic properties of stoichiometric EuO in the temperature range $ T_{P} = $\,\SIrange{350}{600}{\degreeCelsius}. By changing the growth temperature, the total thickness of the EuO film can be varied from \SIrange[range-phrase=--]{9}{25}{nm} for $ t = \SI{60}{\minute} $, and by stopping the growth earlier the thickness can be varied freely. Thus the redox-driven EuO growth method allows to prepare thicker films compared to a topotractic synthesis mechanism. All redox-grown EuO films grown in the suitable parameter window exhibit bulk ferromagnetic properties with no metal inclusions. With regard to the structural properties, we have observed flat and well oriented films with the epitaxial relationship of {EuO(110)$ \parallel $\ce{SrTiO3}(100)}. 

Finally, we demonstrate  that the redox-driven EuO growth scheme can be successfully applied not only to \ce{SrTiO3}, but also to other oxide substrates. We believe, that the universality of a redox-controlled oxide thin film synthesis may also open up exciting perspectives for other topical oxide materials and their integration into complex oxide heterostructures.

\begin{acknowledgments}
We thank David~N.~Müller for fruitful discussions about solid state oxide chemistry. We thank O. Petracic and the Jülich Centre for Neutron science for providing measurement time at the magnetometers.
M.~M. acknowledges financial support from HGF under contract No. VH-NG-811.
\end{acknowledgments}


\bibliography{Library}

\end{document}